\documentclass[pra,reprint,twocolumn,amsmath,nofootinbib, amssymb,superscriptaddress]{revtex4-2}
\usepackage{color}

\usepackage{graphicx}
\usepackage{epsfig}
\usepackage{bm}
\usepackage{soul}
\usepackage[normalem]{ulem}

\newcommand{\bra}[1]{\ensuremath{\left\langle{#1}\right\vert}}

\newcommand{\ket}[1]{\ensuremath{\left|{#1}\right\rangle}}

\def\bea{\begin{eqnarray}}
\def\eea{\end{eqnarray}}

\begin{document}

\title{Chirality-Induced Orbital Selectivity through Linear-Orbital Coupling}

\author{Namgee Cho}
\affiliation{Institut f\"ur Theoretische Physik, Albert-Einstein-Allee 11, Universit\"at Ulm, D-89081 Ulm, Germany}
\author{James Lim}
\affiliation{Institut f\"ur Theoretische Physik, Albert-Einstein-Allee 11, Universit\"at Ulm, D-89081 Ulm, Germany}
\author{Martin B. Plenio}\email{martin.plenio@uni-ulm.de}
\affiliation{Institut f\"ur Theoretische Physik, Albert-Einstein-Allee 11, Universit\"at Ulm, D-89081 Ulm, Germany}

\begin{abstract}
We present a three-dimensional continuum model of electron transmission through a chiral electrostatic potential and show that it gives rise to chirality-induced orbital selectivity. In this model, electron transmittance depends strongly on the incident orbital angular momentum (OAM) associated with its transverse motion, and the selectivity reverses when the potential’s handedness is inverted. The effect originates from a coupling between axial linear momentum and OAM mediated by the helical spatial dependence of the potential. For DNA-scale geometric parameters, this linear-orbital coupling produces sizable orbital selectivity, which remains robust to static and dynamic disorder, and increases with the length of chiral regions. Although bare spin-orbit coupling in the chiral potential considered here is too weak to generate considerable spin dynamics, spin-OAM correlations in the electrodes allow the same orbital selective mechanism to induce appreciable spin selectivity. These results identify orbital dynamics as an important contributor to electron transport in chiral systems.
\end{abstract}
\maketitle

The transfer of electrons through chiral molecules is known to exhibit a pronounced dependence on molecular handedness, exemplified by phenomena such as the chirality-induced spin selectivity (CISS)~\cite{EversAM2022,BloomCR2024}. This effect manifests as substantial enantiospecific differences in electron transmittance, observed in photoemission~\cite{CarmeliAC2002,GohlerScience2011,MollersIJC2022,MollersJPCL2024}, magnetoresistance~\cite{XieNL2011,LuSA2019,MishraJACS2019,JiaACSN2020,NguyenCS2024}, and magnetic resonance experiments~\cite{EckvahlScience2023,EckvahlJACS2024,SantosACSN2018}. Despite extensive investigation, the precise microscopic mechanisms underlying CISS remain elusive, as models based solely on spin-orbit coupling (SOC) fail to fully account for the magnitude of the experimentally measured effects~\cite{EversAM2022}. Theoretical efforts have predominantly relied on simplified electron descriptions, such as low-dimensional tight-binding~\cite{DasJPCC2022,GutierrezPRB2012,GutierrezJPCC2013,GerstenJCP2013,VarelaPRB2016,GeyerJPCC2019,FranssonJPCL2019,Sierra2020,ZhangPRB2020,DuPRB2020,FranssonPRB2020,FranssonNL2021,AlwanJACS2021,LiuNM2021,SmorkaJCP2025,OtsutoPRB2021,AlwanJPCC2024} or one-dimensional continuous-variable models~\cite{VittmannJPCL2022,VittmannJPCL2023,MedinaJCP2015}, due to the complexity inherent in fully three-dimensional simulations. These reduced models have incorporated various physical factors, such as electron-electron correlations~\cite{FranssonJPCL2019,ChiesaNL2024,FranssonNL2021}, spin-phonon coupling~\cite{ZhangPRB2020,DuPRB2020,FranssonPRB2020,FranssonNL2021,SmorkaJCP2025,VittmannJPCL2023,DasJPCC2022,RudgeJCP2025}, and interfacial effects~\cite{AlwanJACS2021,VittmannJPCL2022,AlwanJPCC2024,FranssonNL2021, GhoshJCPL2020}, which have been shown to enhance chirality-dependent electron dynamics. Alternatively, it has been proposed that a substrate with strong SOC generates spin-OAM-correlated electrons, and that the chiral molecule may subsequently filter their OAM states, indirectly inducing spin selectivity via the correlations~\cite{GerstenJCP2013,LiuNM2021}. The potential relevance of OAM in photoelectron scattering has also been studied~\cite{ChenJCTC2026}. However, although previous theoretical studies have suggested the potential importance of orbital angular momentum in CISS, a microscopic mechanism underlying orbital selectivity has not yet been proposed, and the robustness of orbital selectivity under various physical conditions, including variations in the electronic potential landscape and the presence of disorder in chiral systems, remains unexplored.

In this work, we investigate a three-dimensional (3D) continuous-variable model of an electron propagating through a chiral electrostatic potential. The transverse confinement defines local orbital-angular-momentum states around the helical centerline, and transmission through the chiral region depends strongly on the incident OAM state.
Moreover, the orbital polarization changes sign when the handedness of the chiral potential is inverted. In our model, chirality-induced orbital selectivity originates from a helical coupling between axial linear momentum and OAM of an electron, rather than OAM filtering or anisotropic hopping through atomic orbitals in a reduced-dimensional description as in Refs.~\cite{GerstenJCP2013,LiuNM2021} (see the SI for details). 
Our CIOS mechanism is analogous in spirit to CISS induced by delocalized phonon modes~\cite{VittmannJPCL2023}. However, crucially, the linear-orbital coupling, which originates from the transverse electron motion and is responsible for the CIOS effect, is significantly stronger than the spin-phonon coupling under physically reasonable parameter regimes. We demonstrate that the CIOS effect increases with the length of the chiral region and remains robust against static disorder in its internal potential structure. We remark that while bare spin-orbit coupling inside the chiral potential is too weak to induce appreciable spin dynamics in our 3D model, spin-OAM correlations, present prior to transmission through the chiral potential, allow the CIOS mechanism to induce the CISS effect. These results suggest that CIOS may play an important role in enantiospecific electron transmission in chiral systems.

\begin{figure}[t]
    \centering
    \includegraphics[width=1.0\linewidth]{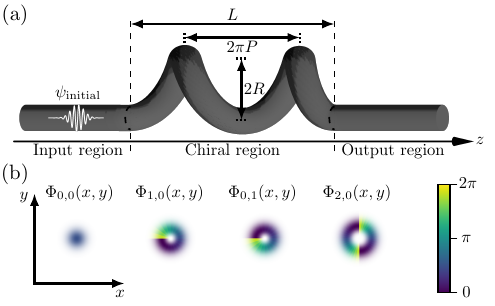}
    \caption{(a) Schematic representation of a 3D model for electron transfer through input-chiral-output regions. The input and output regions are achiral and elongated along the $z$-axis, and are continuously connected to the chiral region, which is parameterized by radius $R$, pitch $2\pi P$, and length $L$. For each $z$, a 2D harmonic potential in the $xy$-plane is considered. (b) Phase distributions of OAM states $\Phi_{n_c,n_d}(x,y)$ in the $xy$-plane. The amplitudes of the OAM states are used as an opacity factor, so that the phase is visualized only where the amplitudes are sufficiently large.}
    \label{fig1}
\end{figure}

We consider a model in three spatial dimensions of an electron propagating through a chiral potential whose long axis coincides with the $z$-axis and is defined in the region $0 \le z \le L$. The Hamiltonian is given by
\begin{align}
    H &= \frac{1}{2m_e}(p_x^2+p_y^2+p_z^2)+\frac{m_e\omega^2}{2}\left(x-R\cos\left(\frac{z}{P}\right)\right)^2 \nonumber\\
    &\quad\, + \frac{m_e\omega^2}{2}\left(y-R\sin\left(\frac{z}{P}\right)\right)^2,\label{eq:H_lab}
\end{align}
where the equilibrium position of the harmonic potential is displaced in the $xy$-plane as a function of $z$, encoding the chirality of the chiral potential, as schematically shown in Fig.~\ref{fig1}(a). To reduce the computational cost of simulating our 3D model, we consider a polaron transformation implemented via a unitary displacement operator conditioned on $z$
\begin{align}
    U = {\rm e}^{iR'{\rm e}^{-iz/P}(c^\dagger - d)+iR'{\rm e}^{iz/P}(c - d^\dagger)},\label{eq:polaron_transform}
\end{align}
with $R'=\sqrt{m_e \omega/4\hbar}R$, where we introduce a set of independent bosonic operators defined as $c = \sqrt{m_e\omega/4\hbar}(y+ix+(ip_y-p_x)/m_e\omega)$, and $d = \sqrt{m_e\omega/4\hbar}(y-ix+(ip_y+p_x)/m_e\omega)$, satisfying the canonical commutation relations $[c,c^{\dagger}]=[d,d^{\dagger}]=1$ and $[c,d^{\dagger}]=0$ (see the SI). Then $H'= U^{\dagger} H U$ is given by
\begin{align}
    H' &= \hbar\omega (c^\dagger c + d^\dagger d+1)\label{eq:H_cd}\\
    &\quad\,+\frac{\left(p_z+i\beta({\rm e}^{iz/P}(c-d^\dagger)-{\rm e}^{-iz/P}(c^\dagger-d))\right)^2}{2m_e}, \nonumber
\end{align}
with $\beta=(R/2P)\sqrt{\hbar m_e \omega}$, where the kinetic energy along the $z$-direction acquires a chiral contribution. In this work, we consider the parameters of DNA, specifically, the radius $R = 1.0\,{\rm nm}$ and the pitch $ 2\pi P = 3.4\,{\rm nm} $, along with the free-electron mass $m_e$. The frequency $\omega$ of the harmonic potential is chosen such that $\hbar\omega \in [0.05,5]\,\rm{eV}$, ensuring that the size of the ground-state wavepacket in the $xy$-plane defined by its standard deviation satisfies $\sqrt{\hbar/(2m_e \omega)}\in [1,10]\,\rm{\text{Å}}$. These parameters lead to $\langle p_z\rangle\beta/2m_e \in [0.15,1.5]\,\rm{eV}$ and $\beta^2/2m_e \in [0.02,2]\,\rm{eV}$ for a reference value of $\langle p_z^2\rangle/2m_e=1\,{\rm eV}$.

In the polaron picture, the $z$-component of the OAM operator is defined as $L_z = x p_y - y p_x=\hbar(c^\dagger c-d^\dagger d)$, where the eigenstates of $L_z$ with OAM $\hbar(n_c-n_d)$ are described by the composite eigenstates $\ket{n_c,n_d}$ of the harmonic modes $c$ and $d$, where $n_c$ and $n_d$ are non-negative integers. Here, positive and negative OAM correspond, respectively, to right- and left-circular motion of the electron about the $z$-axis. In the original frame without the polaron transformation, the $L_z$ operator describes the OAM of an electron with respect to a local coordinate system whose origin follows the chiral path $(x,y)=(R\cos(z/P),R\sin(z/P))$ parameterized by $z$. In Fig.~\ref{fig1}(b), the phase distributions of the OAM states in the $xy$-plane at a fixed $z$ are shown (i.e., $\Phi_{n_c,n_d}(x,y)=\langle x,y|n_c,n_d\rangle$ in the position basis $\ket{x,y}$), demonstrating that the $\Phi_{1,0}$ and $\Phi_{0,1}$ states, which carry non-zero OAM of $\hbar$ and $-\hbar$, respectively, exhibit opposite chiralities in their phase distributions, in contrast to the $\Phi_{0,0}$ state with zero OAM, which shows an achiral phase distribution.

We note that the chiral term in the kinetic energy induces the interaction between linear and orbital angular momenta, such that when the OAM decreases by $\hbar$ via $c-d^\dagger$, the linear momentum increases by $\hbar/P$ via ${\rm e}^{iz/P}$, and vice versa. To demonstrate that this linear-orbital coupling can induce transmittance depending on the initial OAM states, we consider achiral input and output regions in the laboratory frame where the equilibrium position of the harmonic potential in the $xy$-plane is independent of $z$. These regions are smoothly connected to the chiral potential in Eq.~(\ref{eq:H_lab}), as shown in Fig.~\ref{fig1}(a), and serve as a simple model of electrodes coupled to a chiral molecule~\cite{XieNL2011,LuSA2019,MishraJACS2019,JiaACSN2020,NguyenCS2024}. The Hamiltonian for the input and output regions in the polaron picture is given by Eq.~(\ref{eq:H_cd}) with $\beta = 0$. As shown in Fig.~\ref{fig1}(a), we consider an initial wave packet localized in the input region, i.e., $\psi_{\rm initial}\propto \Phi_{n_c,n_d}(x,y){\rm e}^{-(z-z_0)^2/(2\Delta_z^2)+ik_0 z}$, carrying linear momentum $\hbar k_0$ and OAM $\hbar(n_c-n_d)$ about the $z$-axis. We set the initial kinetic energy to ${\rm KE}_0 = (\hbar k_0)^2/2 m_e=1\,{\rm eV}$, and the width to $\Delta_z=4\,{\rm nm}$. The propagation of the electron wave packet is numerically simulated using the finite-difference method.

\begin{figure}[t]
    \centering
    \includegraphics[width=1.0\linewidth]{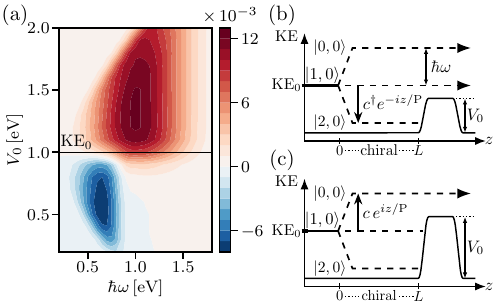}
    \caption{(a) Transmittance difference $\Delta T=T_{1,0}-T_{0,1}$ between the initial OAM states $\ket{1,0}$ and $\ket{0,1}$ as a function of the barrier height $V_0$ and the frequency $\hbar\omega$ of the transverse harmonic potential in the weak-coupling regime ($\beta = \beta_{\rm DNA}/100$). The initial kinetic energy (KE) in the $z$-direction, ${\rm KE}_0 = 1\,{\rm eV}$, is indicated by the solid line. The length of the chiral region is taken as $L=2\pi P$. For the initial state $\ket{1,0}$: (b) Schematic of the effect of transitions between OAM states when $V_0<\rm{KE_0}$. The transition $\ket{1,0}\rightarrow \ket{2,0}$, induced by a linear-orbital coupling $\propto c^\dagger {\rm e}^{-iz/P}$, reduces the kinetic energy along $z$ and causes reflection at a barrier, which results in a lower transmittance than the other initial state $\ket{0,1}$ ($\Delta T<0$). (c) Schematic of the effect of transitions when $V_0>\rm{KE_0}$. The transition $\ket{1,0}\rightarrow \ket{0,0}$, induced by another linear-orbital coupling $\propto c {\rm e}^{iz/P}$, enables electron transmission through a barrier, which results in a higher transmittance than the other initial state $\ket{0,1}$ ($\Delta T>0$).}
    \label{fig2}
\end{figure}

\begin{figure}[t]
    \centering
    \includegraphics[width=1.0\linewidth]{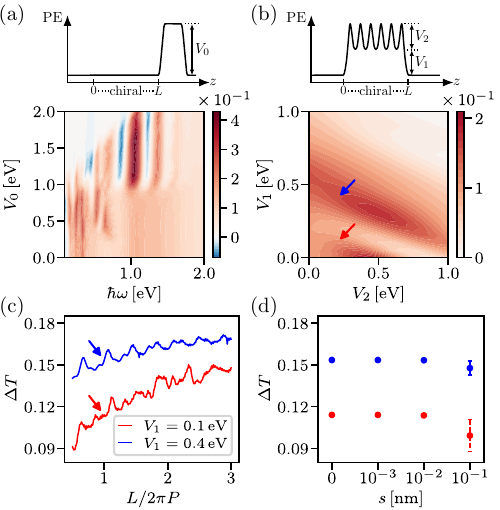}
    \caption{(a) $\Delta T$ as a function of $V_0$ and $\omega$ in the strong-coupling regime ($\beta = \beta_{\rm DNA}$), with $L=2\pi P$, as in Fig.~\ref{fig2}(a). (b) $\Delta T$ as a function of $V_1$ and $V_2$ in the lattice-potential model, with $\beta = \beta_{\rm DNA}$ and $L=2\pi P$. (c) $\Delta T$ as a function of $L$ for two cases, $(V_1,V_2) = (0.1,0.2)\,{\rm eV}$ and $(0.4,0.2)\,{\rm eV}$, marked by arrows in (b). (d) $\Delta T$ as a function of the standard deviation $s$ of static disorder in the lattice-potential structure, with $L=2\pi P$ and $\beta = \beta_{\rm DNA}$, for the two cases marked in (c). Corresponding orbital polarizations are shown in Fig.~SI.4.}
    \label{fig3}
\end{figure}

We proceed to show that electron scattering within the chiral region is governed by (i) energy conservation and (ii) a coupled change in linear and orbital angular momenta which, together, underpin the CIOS effect. In Fig.~\ref{fig2}, we consider two distinct initial states with $(n_c,n_d)=(1,0)$ or $(0,1)$, both having the same linear momentum $\hbar k_0$ in the $z$-direction. To clarify the mechanism of chirality-induced orbital selectivity, we examine a weak-coupling regime in which the chiral coupling strength $\beta$ based on the DNA parameters is reduced by two orders of magnitude (i.e., $\beta=\beta_{\rm DNA}/100$); simulation results using the full chiral coupling strength will be presented in Fig.~\ref{fig3}. In the weak-coupling regime, the electron energy is approximately given by $E\approx \hbar\omega(n_c+n_d+1)+(\hbar k_0)^{2}/2m_e$, implying that the two initial states with opposite OAM have the same energy. Starting from the initial OAM state $\ket{1,0}$, a transition to $\ket{0,0}$ (or $\ket{2,0}$) results in an increase (or decrease) of the linear momentum from $\hbar k_0$ to $\hbar k_1$ due to energy conservation, i.e., $(\hbar^2/2m_e)(k_1^2-k_0^2) = \hbar\omega$ (or $(\hbar^2/2m_e)(k_1^2-k_0^2) = -\hbar\omega$). This transition is mediated by the chiral coupling term proportional to $c {\rm e}^{iz/P}$ (or $c^\dagger {\rm e}^{-iz/P}$), which increases (or decreases) the linear momentum by $\hbar/P$, i.e., $k_1-k_0=1/P$ (or $k_1-k_0=-1/P$). For the initial kinetic energy of ${\rm KE}_0=(\hbar k_0)^{2}/2m_e=1\,{\rm eV}$ and the pitch of $2\pi P=3.4\,{\rm nm}$, both conditions are fulfilled when $\hbar\omega \approx 0.85\,{\rm eV}$ (or $0.59\,{\rm eV}$). For the other initial OAM state $\ket{0,1}$, the two conditions are satisfied only for the transition $\ket{0,1}\rightarrow\ket{1,1}$. Since this transition is induced by the chiral coupling $\propto \bra{1}c^\dagger\ket{0}$, which is weaker than the coupling $\propto \bra{2}c^\dagger\ket{1}$ responsible for the transition $\ket{1,0}\rightarrow \ket{2,0}$, the latter transition occurs with a higher probability. In contrast, transitions induced by the other chiral coupling term $d {\rm e}^{-iz/P}$ (or $d^\dagger {\rm e}^{iz/P}$), such as $\ket{0,1}\rightarrow \ket{0,0}$ (or $\ket{0,1}\rightarrow \ket{0,2}$), cannot simultaneously satisfy energy and momentum conservation because the chiral nature of the coupling modifies the linear momentum in a way that violates energy conservation. Therefore, electron dynamics within the chiral region depends on the initial OAM state.

To demonstrate that different electron dynamics, depending on the initial OAM state, can result in distinct transmission probabilities through the chiral region, in Fig.~\ref{fig2}, we consider a rectangular potential barrier along the $z$-direction, located after the chiral region, with a controlled barrier height $V_0$ and a fixed width of $1\,{\rm nm}$. Even if the barrier is placed inside the chiral region, the transmission remains dependent on the initial OAM state (not shown). We examine the difference in transmittance of the two initial states, i.e., $\Delta T = T_{1,0}-T_{0,1}$, as a function of the barrier height $V_0$ and the frequency $\omega$ of the harmonic potential in the $xy$-plane, where $T_{n_c,n_d}$ denotes the total population of the electron wave packet in the output region after scattering through the chiral region, given the initial OAM state $\ket{n_c,n_d}$. For barrier heights lower than the initial kinetic energy, i.e., $V_0<1.0\,{\rm eV}$, a negative transmittance difference ($\Delta T<0$) appears around $\hbar\omega \approx 0.65\,{\rm eV}$, because the transition from $\ket{1,0}$ to $\ket{2,0}$ lowers the kinetic energy, causing the electron to be reflected by the barrier, and thus reducing the transmittance of the initial OAM state $\ket{1,0}$, as illustrated in Fig.~\ref{fig2}(b). The transition from the other initial state $\ket{0,1}$ to $\ket{1,1}$ likewise results in reflection by the barrier, but occurs with a lower probability. These transient electronic dynamics can be tracked in simulations (see Fig.~SI.1 in the SI). For barrier heights higher than the initial kinetic energy, i.e., $V_0>1.0\,{\rm eV}$, the transmittance difference exhibits a broad positive peak ($\Delta T>0$) centered around $\hbar\omega\approx 1\,{\rm eV}$. In this regime, the initial kinetic energy is insufficient for direct transmission, making the transition from $\ket{1,0}$ to $\ket{0,0}$ crucial, as schematically shown in Fig.~\ref{fig2}(c). We note that the extrema in Fig.~\ref{fig2} occur close to, but not exactly at, the values expected from the bare matching condition, namely $\hbar\omega \approx 0.85\,{\rm eV}$ (or $\hbar\omega \approx 0.59\,{\rm eV}$). This deviation reflects the $\omega$-dependence of the coupling amplitude $\beta = (R/2P)\sqrt{\hbar m_e \omega}$. We verified this by evaluating the transmittance difference for a fixed $\beta$ at a reference frequency $\omega_0$ with $\hbar\omega_0=1\,{\rm eV}$. This yields a positive peak centered at $\hbar\omega \approx 0.85\,{\rm eV}$, and a negative peak at $\hbar\omega\approx 0.59\,\rm{eV}$ (See Fig.~SI.2 in the SI). These results indicate that the coupling strength $\beta$ and the associated transition matrix element, which increase with $\omega$, are responsible for shifting the peak from $0.85\,{\rm eV}$ to $1.0\,{\rm eV}$ (or from $0.59\,{\rm eV}$ to $0.65\,{\rm eV}$).

In Fig.~\ref{fig3}(a), the transmittance difference $\Delta T$ is shown as a function of $V_0$ and $\omega$, similar to Fig.~\ref{fig2}(a), but now for the linear-orbital coupling based on DNA-scale parameters ($\beta = \beta_{\rm DNA}$). In this strong-coupling regime, the electron’s energy can no longer be expressed simply as the sum of the OAM energy and the kinetic energy in the $z$-direction, and thus the analysis used in the weak-coupling regime of Fig.~\ref{fig2} is no longer applicable. However, the strong coupling gives rise to richer OAM dynamics and enhanced orbital selectivity. In contrast to the weak-coupling case, orbital selectivity with non-zero $\Delta T$ appears even in the absence of the potential barrier ($V_0 = 0$), due to the OAM-dependent reflection at the interface between the input and chiral regions (see Fig.~SI.3 in the SI). Notably, $\Delta T$ in Fig.~\ref{fig3}(a) is one to two orders of magnitude larger than in the weak-coupling case in Fig.~\ref{fig2}(a), over a broad range of $V_0$ and $\omega$. This enhanced orbital selectivity still originates from the coupled change in linear and orbital angular momenta, as can be seen from the fact that $\Delta T$ vanishes when the factors ${\rm e}^{\pm iz/P}$ in Eq.~(\ref{eq:H_cd}), which shift the linear momentum conditioned on the OAM transitions, are removed. We find that for $V_0\lesssim{\rm KE}_0 = 1\,{\rm eV}$, $\Delta T$ remains positive, whereas for $V_0\gtrsim 1\,{\rm eV}$, $\Delta T$ can take both positive and negative values. For the data shown in Fig.~\ref{fig3}(a), the orbital polarization, defined as the transmittance difference $\Delta T$ divided by the total transmission probability, reaches several tens of percent, up to $80\,\%$ (see Fig.~SI.4 in the SI). This high orbital polarization arises primarily from the initial-OAM-dependent transition to $\ket{0,0}$, analogous to the weak-coupling case shown in Fig.~\ref{fig2}(c), as confirmed numerically by the OAM spectrum of the transmitted wave packet (see Fig.~SI.5 in the SI). When the handedness of the 3D chiral potential is reversed, the sign of $\Delta T$ and the orbital polarization also reverse, while the line shapes in Fig.~\ref{fig3}(a) remain unchanged.

So far, we have not considered any internal potential structure within the chiral region. We now introduce a lattice-like potential as a function of $z$ inside the chiral region, parameterized by an offset $V_1$ and the height $V_2$ of multiple Gaussian potentials, as schematically illustrated in Fig.~\ref{fig3}(b), with fixed $\hbar\omega=0.5\,\rm{eV}$. For a single chiral turn with $L = 2\pi P$, we place six Gaussian potentials with uniform spacing and each of the form ${\rm e}^{-(z - z_i)^2 / 2 \delta^2}$, centered at $z = z_i$, with width $\delta = P/5$. As shown in Fig.~\ref{fig3}(b), positive $\Delta T$ on the order of $10^{-1}$ emerges over a broad range of $V_1$ and $V_2$. For the two representative points indicated by arrows in Fig.~\ref{fig3}(b), the dependence of $\Delta T$ on the length $L$ of the chiral region is shown in Fig.~\ref{fig3}(c). Here, $L$ increases continuously, so the final Gaussian potential at the boundary between the chiral and output regions may not be fully contained within the chiral region. In the simulations, this final Gaussian potential is smoothly suppressed using a smooth step function to ensure a continuous transition into the output region. Notably, $\Delta T$ gradually increases with the number of chiral turns, from one to three. This length dependence is associated with the presence of the multiple Gaussian potentials, as $\Delta T$ shows negligible length dependence when $V_2 = 0$ (see Fig.~SI.6 in the SI). These results indicate that the orbital selectivity arises not only at the interfaces between the chiral and input/output regions, but also within the chiral region itself, and that this length dependence occurs even in the absence of electronic dephasing noise. 

In Fig.~\ref{fig3}(d), we examine the effect of static disorder in the lattice potential structure for the cases marked by arrows in Fig.~\ref{fig3}(c), with $L = 2\pi P$. Starting from the positions of the six Gaussian potentials with uniform spacing, as considered in Figs.~\ref{fig3}(b) and (c), the position of each Gaussian is independently shifted, with each shift randomly generated from a zero-mean Gaussian distribution with a controlled standard deviation $s$. As shown in Fig.~\ref{fig3}(d), the magnitude of $\Delta T$ remains essentially unchanged for $s$ up to $10^{-2}\,{\rm nm}$ and shows only a minor change at $10^{-1}\,{\rm nm}$, covering a broad range of molecular deformation scales induced by phonon motion. Additionally, the CIOS effect remains robust under dynamic disorder, in which the amplitudes of the Gaussian potentials are time-dependent and the electron energy is consequently not conserved (see the SI). These results demonstrate that the orbital selectivity observed in our work is robust against structural disorder in the lattice potential of the chiral region.

In this work, we have demonstrated that a 3D chiral potential with physically reasonable parameters can induce significant orbital selectivity, where the electron transmittance depends on the initial orbital angular momentum state. While the model in the main text assumes a harmonic confinement potential, qualitatively similar results are obtained for anharmonic confinement (see the SI), highlighting the generality of the CIOS effect. Similar CIOS effects are also observed in tight-binding models based on atomic orbitals, and these results will be reported in a forthcoming paper. By employing a wave-packet method, which enables monitoring of transient electron dynamics during scattering inside the chiral region, we have clarified a CIOS mechanism in terms of energy and momentum conservation. While we have considered two specific initial OAM states, $\ket{1,0}$ and $\ket{0,1}$, and examined their difference in transmittance in the main text, similar CIOS effects are observed for other initial OAM states (see Fig.~SI.9 in the SI).

For the parameters considered in Figs.~\ref{fig2} and \ref{fig3}, we found that the bare spin-orbit coupling $H_{\rm SOC}=\hbar(2m_e c)^{-2}\boldsymbol{\sigma}\cdot(\nabla W\times \mathbf{p})$ induced by the chiral potential energy $W$ in Eq.~(\ref{eq:H_lab}) results in negligible spin dynamics, where $\boldsymbol{\sigma}$ and $\mathbf{p}$ denote the Pauli spin operator and the electron's momentum operator, respectively. However, when the electron's OAM and spin states are correlated prior to transmission through the chiral region, orbital selectivity can give rise to spin selectivity. In the SI, we introduce spin-orbit interaction at the interface between the input and chiral regions to account for the finite SOC of the electrodes. When electrons with random spin and zero OAM enter the nonchiral SOC region, spin flips are accompanied by changes in OAM to conserve the total angular momentum about the long axis of the input region (i.e., $\ket{0,0,\uparrow} \rightarrow \ket{1,0,\downarrow}$ and $\ket{0,0,\downarrow} \rightarrow \ket{0,1,\uparrow}$). Since OAM states exhibit different transmission probabilities through the chiral region, the spin-OAM correlations lead to both spin and orbital selectivity (see the SI for details).

The chiral terms in Eq.~(\ref{eq:H_lab}), i.e., those proportional to $x\cos(z/P)$ and $y\sin(z/P)$, are analogous to the spin-phonon coupling of Ref.~\cite{VittmannJPCL2023}, originating from fluctuations of the spin-orbit coupling of a chiral molecule induced by delocalized phonon motion. The spin-phonon interaction of Ref.~\cite{VittmannJPCL2023} was modeled by $\sigma_{\rm SOC} A \sin(2\pi s/\lambda)$ where $\sigma_{\rm SOC}$ is a spin operator, $A$ is the amplitude of a harmonic phonon mode with wavelength $\lambda$, and $s$ is a coordinate describing the electron's motion along a chiral one-dimensional path. However, there are two key differences. First, in the spin-phonon coupling model, spin selectivity is induced by delocalized phonon modes that typically have low vibrational frequencies ($\lesssim 0.01\,{\rm eV}$) that are one to two orders of magnitude smaller than the energy quanta $\hbar\omega$ associated with orbital angular momentum states. Such low phonon frequencies may lead to electron transmittance that is less sensitive to the barrier height than the orbital selectivity present in this work. Second, the spin-phonon coupling strength is expected to be weaker than the bare spin-orbit interaction, as the spin-phonon coupling arises as a perturbation of the spin-orbit coupling, suggesting that it is likely to be weak in realistic systems. In contrast, the linear-orbital coupling $\beta$, which induces orbital selectivity, has a notable magnitude under physically reasonable parameters. This makes orbital selectivity potentially more relevant and effective than the spin-phonon mechanism in influencing electron transport in chiral systems.

To clarify the relevance of the CIOS mechanism in CISS experiments, a detailed microscopic description of both the chiral molecules and the electrodes is required, for example using tight-binding models with parameters derived from first-principles calculations. Such a framework may enable quantitative comparison with experimental observations of CISS effects and thereby provide a way to assess the viability of the CIOS mechanism. More direct verification would require experimental schemes capable of resolving the OAM states of electrons transmitted through chiral molecules. One possible route is via photoelectron-based experiments. For example, one could analyze how electron OAM and spin polarizations contribute to the diffraction patterns produced by a grating, for which OAM-dependent diffraction has been demonstrated for electron vortex beams~\cite{GuzzinatiPRA2014,SaitohPRL2013,GrilloNC2017}, or to the scattering patterns measured by Mott polarimetry~\cite{GohlerScience2011,MollersIJC2022,BoxemPRA2014,ChenJCTC2026}.

Our results demonstrate the critical role of orbital angular momentum dynamics and chirality-induced orbital selectivity in enantiospecific electron transmission. These findings complement and extend the recent advances in orbitronics, where chiral materials and crystals are being recognized as promising platforms to generate and control orbital angular momentum for novel transport and device functionalities~\cite{YenNP2024,BrinkmanPRL2024,JoNPJS2024}.

\section*{Acknowledgements}

We thank Thibaut Lacroix and Susana F. Huelga for helpful discussions at the early stages of this project. This work was supported by the ERC Synergy grant HyperQ (Grant No.~856432), the BMBF via project PhoQuant (Grant No.~13N16110), and the Volkswagen Foundation (Grant No.~0200187). The authors acknowledge support by the state of Baden-W{\" u}rttemberg through bwHPC and the German Research Foundation (DFG) through Grant No.~INST 40/575-1 FUGG (JUSTUS 2 cluster).

\bibliographystyle{apsrev4-2}
\bibliography{references_main}
\clearpage

\renewcommand{\thefigure}{SI.\arabic{figure}}
\setcounter{figure}{0}

\renewcommand{\theequation}{SI.\arabic{equation}}
\setcounter{equation}{0}
\begin{widetext}

\begin{center}
\textbf{\large Supporting Information}
\end{center}

\section{Polaron transformation}

In the main text, we consider the Hamiltonian of the form
\begin{align}
    H &= \frac{1}{2m_e}(p_x^2+p_y^2+p_z^2)+\frac{m_e\omega^2}{2}\left(x-R\cos(z/P)\right)^2 + \frac{m_e\omega^2}{2}\left(y-R\sin(z/P)\right)^2,\label{eq:S1}
\end{align}
where the equilibrium position of the harmonic potential is displaced in the $xy$-plane, as a function of $z$. The unitary operator for the polaron transformation, defined in the main text, can be re-expressed as 
\begin{align}
    U = {\rm exp}\left(R''\cos(z/P)(a^\dagger - a)+R''\sin(z/P)(b^\dagger - b)\right),
\end{align}
with $R''=\sqrt{m_e \omega/2\hbar}R$, where $a = \sqrt{m_e\omega/2\hbar}(x+ip_x/m_e\omega)$ and $b = \sqrt{m_e\omega/2\hbar}(y+ip_y/m_e\omega)$ denote, respectively, the annihilation operators of the harmonic oscillators in the $x$- and $y$-directions. The unitary operator converts the Hamiltonian $H$ into
\begin{align}
    H' &= U^{\dagger} H U \\
    &= \frac{1}{2m_e}(p_x^2+p_y^2)+\frac{m_e\omega^2}{2}(x^2+y^2)+\frac{1}{2m_e}\left(p_z+\frac{R}{P}\sin(z/P)p_x-\frac{R}{P}\cos(z/P)p_y \right)^2, \\
    &= \hbar\omega (a^\dagger a + b^\dagger b+1)+\frac{\left(p_z+i\beta'\left(\sin(z/P)(a^\dagger -a)-\cos(z/P)(b^\dagger - b)\right) \right)^2}{2m_e}, 
\end{align}
with $\beta'=(R/P)\sqrt{\hbar m_e \omega/2}$, where the harmonic potential in the $xy$-plane becomes achiral, while the kinetic energy along the $z$-direction acquires a chiral contribution. A similar polaron transformation has been shown to decrease the computational cost of spin-boson models, where the equilibrium positions of harmonic oscillators are shifted depending on spin states. We find that our polaron transformation likewise reduces the computational cost for simulating our 3D model.

In the polaron picture, the OAM operator in the $z$-direction is given by $L_z = x p_y-y p_x=i\hbar (ab^\dagger - a^\dagger b)$. To consider the OAM eigenstates of $L_z$ explicitly, we introduce in the main text a new set of independent bosonic operators defined as $c=(b+i a)/\sqrt{2}$ and $d=(b-i a)/\sqrt{2}$, satisfying the canonical commutation relations $[c,c^{\dagger}]=[d,d^{\dagger}]=1$ and $[c,d^{\dagger}]=0$. The OAM operator is then expressed as $L_z=\hbar(c^\dagger c-d^\dagger d)$. The Hamiltonian $H'$ in the polaron picture can also be expressed in terms of the new mode operators $c$ and $d$
\begin{align}
    H' &= \hbar\omega (c^\dagger c + d^\dagger d+1)+\frac{\left(p_z+i\beta({\rm e}^{iz/P}(c-d^\dagger)-{\rm e}^{-iz/P}(c^\dagger-d))\right)^2}{2m_e},
\end{align}
with $\beta=\beta'/\sqrt{2}$.

\begin{figure*}[b]
    \includegraphics[width=\linewidth]{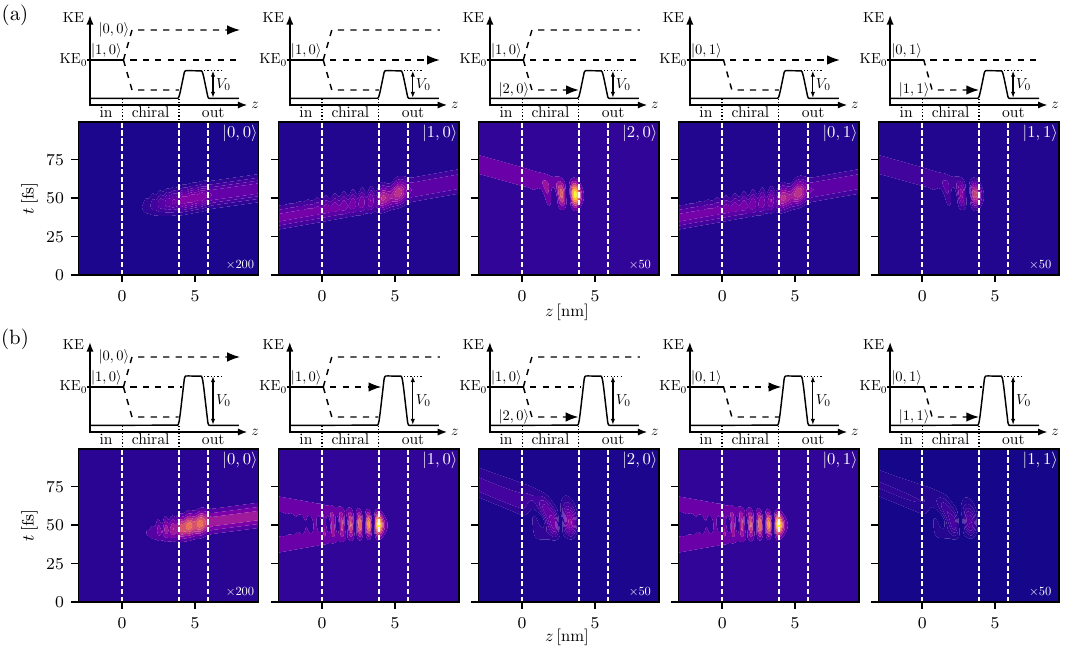}
    \caption{Transient population dynamics of the electron wave packet conditioned on the OAM states $\ket{n_c,n_d}$, as a function of time $t$ and coordinate $z$, for the settings corresponding to those used in Fig.~2 of the main text, where the initial OAM state is either $\ket{1,0}$ (the three leftmost columns) or $\ket{0,1}$ (the remaining columns). Each panel displays the population dynamics of a specific OAM component at position $z$, defined as $|\langle n_c,n_d,z|\psi(t)\rangle|^2$, where $\ket{\psi(t)}$ denotes the full electron state at time $t$, evolved under the Hamiltonian $H'$ in the polaron picture. In (a) and (b), we show the cases $(V_0,\hbar\omega)=(0.7,0.59)\,{\rm eV}$ and $(V_0,\hbar\omega)=(1.2,0.85)\,{\rm eV}$, respectively, which correspond to the minima and maxima of the transmittance difference $\Delta T$ in Fig.~2(a), where the energy and momentum conservation conditions discussed in the main text are satisfied. In simulations, we assume that the $z$-dependence of the initial states is Gaussian; $\psi_{\rm initial}\propto {\rm e}^{-(z-z_0)^2/(2\Delta_z^2) + ik_0z}$, with kinetic energy ${\rm KE}_0 = (\hbar k_0)^2/2m_e=1\,{\rm eV}$ and the width $\Delta_z = 4\,{\rm nm}$. The center position $z_0$ is chosen so that the initial wave packet is well localized in the input region.}
    \label{figS1}
\end{figure*}

\begin{figure}
    \includegraphics[width=0.5\linewidth]{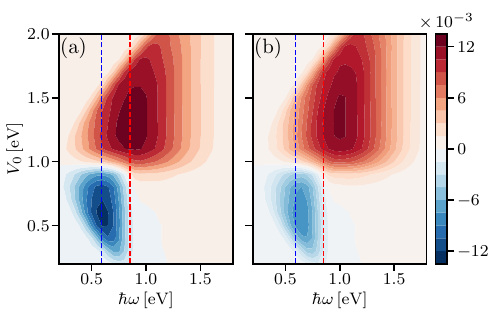}
    \caption{(a) $\Delta T$ when $\beta = (R/2P) \sqrt{\hbar m_e \omega_0}$, with fixed $\hbar\omega_0 = 1\,{\rm eV}$. Vertical dashed lines indicate $\hbar\omega \in \{0.59, 0.85\}\,{\rm eV}$, where a minimum and a maximum in $\Delta T$ are expected to occur according to energy conservation and the linear momentum shift of $\hbar/P$ induced by the chiral coupling, as discussed in the main text. (b) $\Delta T$ when $\beta = (R/2P) \sqrt{\hbar m_e \omega}$, as in Fig.~2(a), demonstrating that the $\omega$-dependence of $\beta$ shifts the positions of the $\Delta T$ extrema.}
    \label{figS2}
\end{figure}

\begin{figure}
    \includegraphics[width=0.5\linewidth]{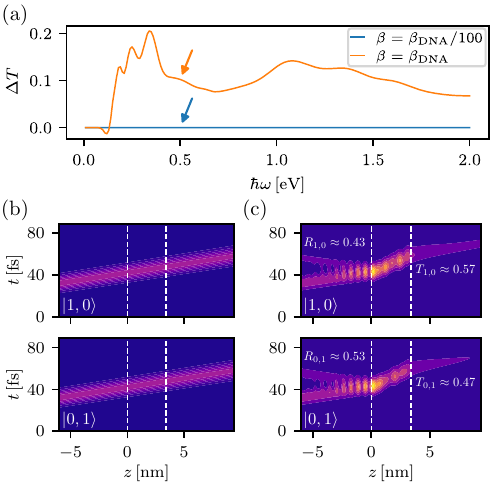}
    \caption{(a) $\Delta T$ in the absence of a potential barrier ($V_0=0$, see Figs.~2 and 3(a) in the main text) for the weak-coupling ($\beta=\beta_{\rm DNA}/100$) and strong-coupling ($\beta=\beta_{\rm DNA}$) regimes. (b,c) The transient dynamics of the electron wave packet are shown as functions of time $t$ and coordinate $z$ for (b) the weak-coupling and (c) the strong-coupling cases, with $\hbar\omega=0.5\,{\rm eV}$ marked by arrows in (a). Only in the strong-coupling case does notable reflection occur at the interface between the input and chiral regions, depending on the initial OAM state $\ket{1,0}$ or $\ket{0,1}$. The reflection probabilities for the initial OAM states, computed at a fixed time immediately after reflection, are $R_{1,0}\approx 0.43$ and $R_{0,1}\approx 0.53$. When summed with the final transmission probabilities $T_{1,0}\approx 0.57$ and $T_{0,1}\approx 0.47$, the results are close to unity, indicating that reflection primarily occurs at the interface between the input and chiral regions.}
    \label{figS3}
\end{figure}

\begin{figure}
    \includegraphics[width=0.5\linewidth]{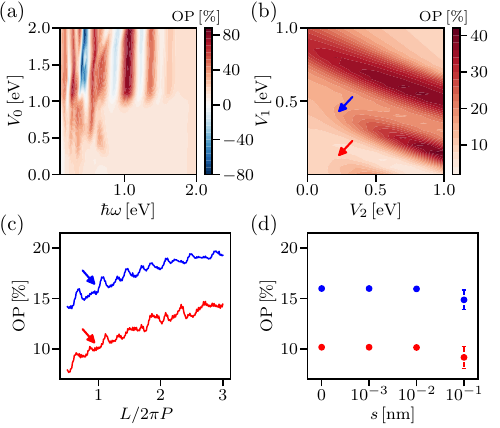}
    \caption{Orbital polarization, defined as ${\rm OP}=(T_{1,0}-T_{0,1})/(T_{1,0}+T_{0,1})$, for the results shown in Fig.~3 of the main text.}
    \label{figS4}
\end{figure}

\begin{figure}
    \includegraphics[width=0.5\linewidth]{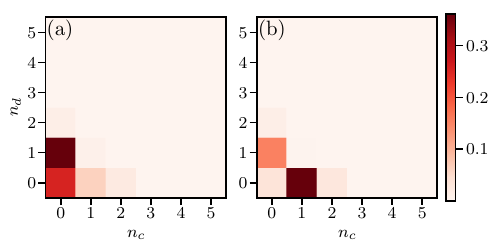}
    \caption{For the initial OAM states (a) $\ket{1,0}$ and (b) $\ket{0,1}$ considered in Fig.~3(a) of the main text, the OAM spectra of the electron wave packets transmitted through the chiral potential are shown. Here, the potential barrier after the chiral region is absent, i.e., $\hbar\omega = 1.0\,\rm{eV}$ and $V_0 = 0$. In (a), a high population of the $\ket{0,0}$ state is observed, indicating that the $\ket{1,0}\rightarrow\ket{0,0}$ transition favorably occurs within the chiral region. In contrast, in (b), the population of the $\ket{0,0}$ state is negligible because the $\ket{0,1}\rightarrow\ket{0,0}$ transition is weak. These OAM spectra demonstrate that the transition to $\ket{0,0}$ is much more probable for the initial state $\ket{1,0}$. When a potential barrier is introduced after the chiral region, the $\ket{1,0}\rightarrow\ket{0,0}$ transition provides sufficient longitudinal kinetic energy for the electron to overcome the barrier, thereby rationalizing the positive transmittance difference, $\Delta T>0$, observed in the simulations for $\hbar\omega = 1.0\,\rm{eV}$ and $V_0 > 1.0\,\rm{eV}$ (see Fig.~3(a) of the main text).}
    \label{figS5}
\end{figure}

\begin{figure*}
    \includegraphics[width=\linewidth]{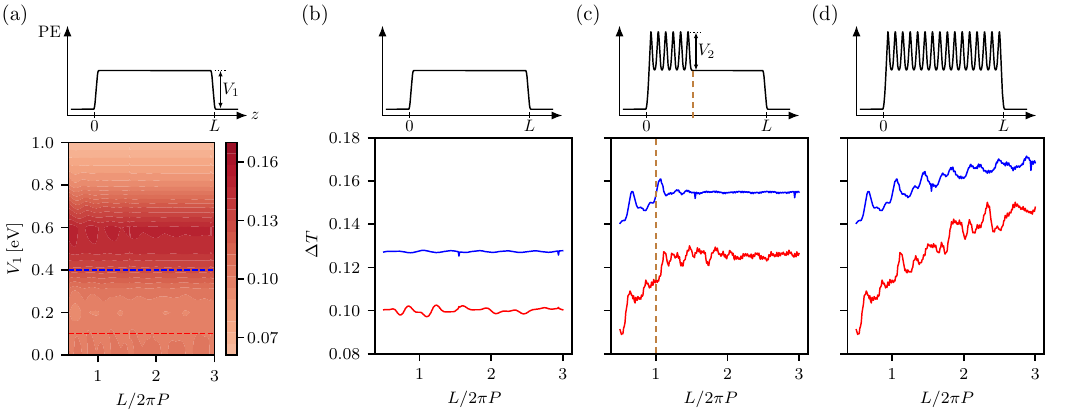}
    \caption{(a) $\Delta T$ as a function of the length $L$ and the potential offset $V_1$ of the chiral region without multiple Gaussian potentials ($V_2=0$, see Figs.~3(b) and (c) in the main text), showing negligible length dependence. (b) $\Delta T$ as a function of $L$ for two representative offsets, $V_1 \in \{0.1,0.4\}\,{\rm eV}$, marked by red and blue dashed lines in (a). (c) $\Delta T$ when multiple Gaussian potentials with $V_2=0.2\,{\rm eV}$ are introduced up to the first chiral turn, producing notable length dependence up to $L \lesssim 2\pi P$ and negligible dependence beyond that. (d) $\Delta T$ when multiple Gaussian potentials with $V_2=0.2\,{\rm eV}$ are present throughout the entire chiral region, resulting in notable length dependence over one to three chiral turns. These results indicate that the length dependence is strongly associated with the presence of multiple Gaussian potentials in the chiral region. We note that electronic dephasing is not required to observe the length dependence in our model.}
    \label{figS6}
\end{figure*}

\begin{figure}
    \includegraphics[width=0.5\linewidth]{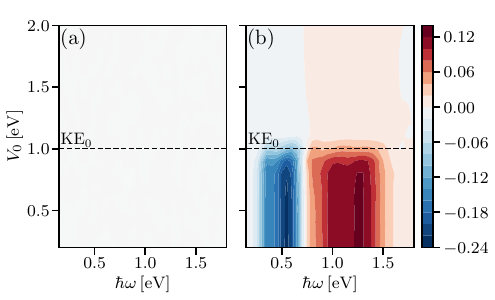}
    \caption{$\Delta T$ when the OAM energy term $\hbar\omega(c^\dagger c + d^\dagger d)$ is omitted in (a) the weak-coupling ($\beta=\beta_{\rm DNA}/100$) and (b) the strong-coupling ($\beta=\beta_{\rm DNA}$) regimes (see Figs.~2 and 3(a) in the main text).}
    \label{figS7}
\end{figure}

\section{Differences from previous CIOS studies}
In the following, we would like to draw attention to and discuss important differences of our model compared to those reported in earlier works, notably Refs.~\cite{GerstenJCP2013,LiuNM2021}.

\noindent
{\em Underlying physical mechanisms --} In Refs.~\cite{GerstenJCP2013,LiuNM2021}, tight-binding models for CIOS were investigated numerically, but the underlying physical mechanisms that give rise to the selectivity were not examined in detail. Instead, these studies primarily focus on the emergence of spin selectivity arising from the interplay of CIOS and the correlations between OAM and spin states induced by strong SOC in a substrate. In contrast, our work identifies the mechanism underlying CIOS as originating from the combination of (i) a coupled change in the OAM and linear momentum along the $z$-direction due to the chirality of the scattering potential, and (ii) the principles of energy and momentum conservation.

{\em Hilbert space and energetic structure --} In Ref.~\cite{LiuNM2021}, a tight-binding model in one spatial dimension with three sites per chiral turn was studied. Each site was modeled by three atomic $p$-orbitals of identical energy, giving rise to three fully degenerate OAM states with OAM limited to $+\hbar$, 0, and $-\hbar$. In contrast, our 3D model features electron OAM states $\ket{n_c,n_d}$ ($n_c$ and $n_d$ are non-negative integers) that are only partially degenerate, i.e., $\hbar\omega(c^\dagger c+d^\dagger d)\ket{n_c,n_d}=\hbar\omega(n_c+n_d)\ket{n_c,n_d}$, and allow for unbounded OAM $\hbar(n_c-n_d)$. In the presence of strong coupling that induces transitions between OAM states, restricting the OAM Hilbert space to only $+\hbar$, 0, and $-\hbar$ may not accurately capture the electronic dynamics.

Because of energy conservation, the full degeneracy of the OAM states, as assumed in Ref.~\cite{LiuNM2021}, can significantly influence the CIOS effect. In our model, a coupled change in the OAM and linear momentum along the $z$-direction is crucial. Due to the energy differences of OAM states, this may lead to changes in the kinetic energy in the $z$-direction, which in turn affects transmittance through the chiral region containing potential barriers, as detailed in the main text. If the OAM energy term $\hbar\omega(c^\dagger c+d^\dagger d)$ is removed from our simulations, making the OAM states fully degenerate, a change in OAM cannot alter the kinetic energy due to energy conservation.

In Fig.~\ref{figS7}, we consider a rectangular potential barrier located after a chiral region, as in Figs.~2 and 3(a) of the main text, and show how omitting the OAM energy term $\hbar\omega(c^\dagger c+d^\dagger d)$ affects the transmittance difference $\Delta T=T_{1,0}-T_{0,1}$ in simulations.

In the weak-coupling regime ($\beta = \beta_{\rm DNA}/100$), the transmittance difference $\Delta T$ vanishes within numerical accuracy when $\hbar\omega(c^\dagger c+d^\dagger d)$ is ignored, as shown in Fig.~\ref{figS7}(a). The full degeneracy of the OAM states does not allow for changes in the kinetic energy for motion along the $z$-direction, even if transitions between OAM states occur. Because the coupling between linear momentum and OAM involves a change in the kinetic energy, the energy and momentum conservation conditions described in the main text cannot be satisfied simultaneously. As a result, all OAM transitions are suppressed in the weak-coupling regime, leading to $\Delta T \approx 0$.

In the strong-coupling regime ($\beta = \beta_{\rm DNA}$), even if the OAM energy term $\hbar\omega(c^\dagger c+d^\dagger d)$ is disregarded, transitions between OAM states can occur. However, when the OAM states are fully degenerate, the kinetic energy along $z$ may not increase during OAM transitions. In this case, the electron is expected to be almost completely reflected by the potential barrier when its initial kinetic energy, ${\rm KE}_0 = 1\,{\rm eV}$, is lower than the barrier height $V_0$. This prediction is consistent with the simulated results in Fig.~\ref{figS7}(b), where omitting $\hbar\omega(c^\dagger c+d^\dagger d)$ leads to negligible transmission probabilities for $V_0 > 1\,{\rm eV}$, resulting in $\Delta T \approx 0$. This is in contrast to the full model results shown in Fig.~3(a) of the main text, where significant $\Delta T$ is observed for $V_0 > 1\,{\rm eV}$.

{\em Choice of parameters --} Another significant difference between our work and that reported in Refs.~\cite{GerstenJCP2013,LiuNM2021} is that, once the effective three-dimensional potential is specified, the continuum model can be simulated without introducing an additional tight-binding parameterization of hopping amplitudes or onsite orbital energies. The transport behavior is then determined by the geometric, confinement, barrier, and disorder parameters used to define the effective scattering problem.

{\em Electronic dephasing noise --} Our continuous-variable 3D model does not require any electronic dephasing to observe the CIOS effects present in our work. In contrast, tight-binding models often require dephasing noise to observe CIOS/CISS effects.

{\em Access to transient dynamics --} We also note that our work employs a wave-packet method, which enables monitoring of transient electron dynamics at the interfaces between the input/output and chiral regions, as well as within the chiral region itself (see Figs.~\ref{figS1} and \ref{figS3}). This wave-packet approach helps clarify the underlying CIOS mechanisms behind the simulated results and contrasts with the non-equilibrium Green's function method employed in Refs.~\cite{GerstenJCP2013,LiuNM2021}, which computes transmittance but does not provide access to transient dynamics.

\begin{figure}
    \includegraphics[width=0.5\linewidth]{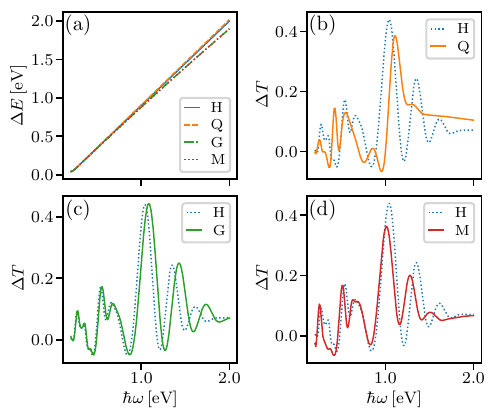}
    \caption{(a) Energy gap $\Delta E$ between the ground and the first excited states of the harmonic (H), quartic (Q), Gaussian (G) and Morse (M) potentials in the $xy$-plane at fixed $z$, shown as a function of $\omega$. (b-d) Transmittance difference for initial states carrying OAM of $\hbar$ or $-\hbar$. See the SI text for further details.}
    \label{figS8}
\end{figure}

\section{Anharmonic Continuous-Variable Models}

In the main text, we consider a continuous-variable 3D model in which the electron's confinement in the transverse $(x,y)$ directions is described by a 2D harmonic potential. Here, we show that the CIOS effects remain robust even when the confinement potentials are anharmonic.

In Fig.~\ref{figS8}, we consider the following anharmonic potentials in the chiral region ($0\le z\le 2\pi P$): 
\begin{align}
    {\rm Quartic}:\quad & W_{Q}(x,y,z) = v_{Q}((x-R\cos(z/P))^4+(y-R\sin(z/P))^4), \label{eq:anharmonic_quadratic}\\
    {\rm Gaussian}:\quad & W_{G}(x,y,z) =v_{G}(1-{\rm e}^{-((x-R\cos(z/P))^2+(y-R\sin(z/P))^2)/2\sigma_{G}^2}), \\
    {\rm Morse}:\quad & W_{M}(x,y,z) = v_{M}((1-{\rm e}^{-\alpha_{M} (x\cos(z/P)+y\sin(z/P)-R)})^2 +(1-{\rm e}^{-\alpha_{M} (-x\sin(z/P)+y\cos(z/P))})^2).\label{eq:anharmonic_morse}
\end{align}
For the input and output regions, we consider the anharmonic potentials whose equilibrium positions are independent of $z$, by fixing the value of $z$ to 0 for $z<0$ and $2\pi P$ for $z>2\pi P$ in Eqs.~(\ref{eq:anharmonic_quadratic})-(\ref{eq:anharmonic_morse}). This results in straight waveguides smoothly connected to the chiral waveguides, as in the harmonic case considered in the main text. We express the parameters of each anharmonic potential as functions of $\omega$ from the harmonic case, namely, $v_{Q}=m_e^2\omega^3/5\hbar$, $v_{G}/\sigma_{G}^2=m_e\omega^2$ and $v_{M}\alpha_{M}^2=m_{e}\omega^2/2$, so that the energy-gap $\Delta E$ between the ground and the first excited states of the anharmonic potential is approximately equal to the energy quanta $\hbar\omega$ of the harmonic case, as shown in Fig.~\ref{figS8}(a).

\begin{figure}
    \includegraphics[width=0.5\linewidth]{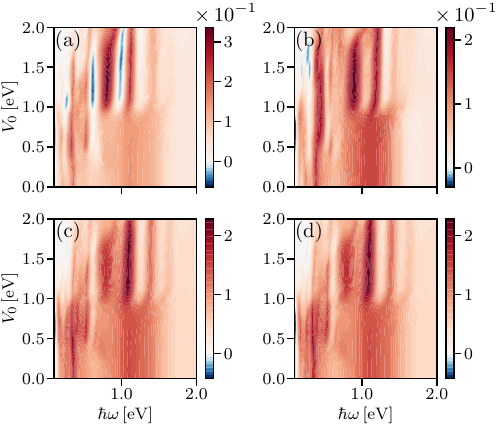}
    \caption{In the main text, the transmittance difference between the two initial OAM states $\ket{1,0}$ and $\ket{0,1}$ is considered. Here, using the same model parameters as in Fig.~3(a) of the main text, we examine additional pairs of initial OAM states and show their transmittance differences: (a) $\ket{2,1}$ and $\ket{1,2}$; (b) $\ket{3,2}$ and $\ket{2,3}$; (c) random mixtures of $\ket{1,0}$, $\ket{2,1}$, and $\ket{3,2}$, and of $\ket{0,1}$, $\ket{1,2}$, and $\ket{2,3}$; and (d) the coherent superpositions $3^{-1/2}(\ket{1,0}+\ket{2,1}+\ket{3,2})$ and $3^{-1/2}(\ket{0,1}+\ket{1,2}+\ket{2,3})$. The initial OAM states $\ket{1,0}$, $\ket{2,1}$, and $\ket{3,2}$ all carry an OAM of $\hbar$ about the $z$-axis, whereas $\ket{0,1}$, $\ket{1,2}$, and $\ket{2,3}$ carry an OAM of $-\hbar$ about the $z$-axis. The initial longitudinal kinetic energy along the $z$-direction is set to $1.0\,{\rm eV}$, independent of the initial OAM state. The transmittance differences shown in (c) and (d) are nearly identical, indicating that coherence between the initial OAM states does not play a significant role in the scattering process within the chiral region.}
    \label{figS9}
\end{figure}

As in Fig.~3(a) of the main text, we consider a rectangular potential barrier located after the chiral region with a fixed potential height of $V_0=1.2\,\rm{eV}$. We assume that the initial state, localized in the input region, is a superposition of the two degenerate first excited states of the anharmonic potential with a relative phase of $i$ or $-i$, similar to $\ket{0,1} \pm i \ket{1,0}$ in the harmonic case, multiplied by a Gaussian wave packet as a function of $z$, carrying the initial kinetic energy of ${\rm KE}_0=1\,{\rm eV}$. These initial states carry OAM of approximately $\hbar$ or $-\hbar$, which is conserved within the input region until the wave packet enters the chiral region. The dissociation energies of the Gaussian and Morse potentials are fixed to $v_{\rm{G}}=v_{\rm{M}}=20\,\rm{eV}$, ensuring that the electron remains confined within the potentials. In Figs.~\ref{figS8}(b)-(d), the transmittance difference between the two initial states is shown, indicating that the orbital selectivity of the harmonic and anharmonic models is qualitatively similar. These results demonstrate that the CIOS effects are robust against variations in the confinement potentials.

\begin{figure}[h!]
    \includegraphics[width=0.5\linewidth]{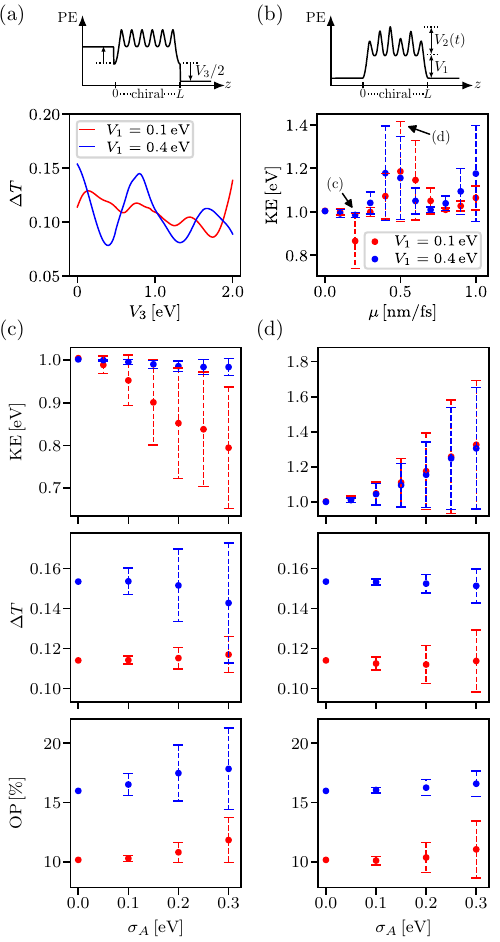}
    \caption{(a) $\Delta T$ as a function of the bias $V_3$, for the time-independent Gaussian potential model from Fig.~3 of the main text, with $V_1\in\{0.1,0.4\}\,{\rm eV}$ and $V_2=0.2\,{\rm eV}$. (b) The kinetic energy of the output electron in the 1D model (no orbital degrees of freedom) as a function of $\mu$ in the dispersion relations of the time-dependent Gaussian potential model (see SI text). For (c) $\mu=0.2\,{\rm nm/fs}$ and (d) $\mu=0.5\,{\rm nm/fs}$, the kinetic energy of the output electron (top panel, 1D model), the transmittance difference $\Delta T$ (middle panel, 3D model), and the orbital polarization (bottom panel, 3D model) are shown.}
    \label{figS10}
\end{figure}

\section{Finite Bias and Time-Dependent Potentials}

In the main text, we consider the zero-bias case, with no potential energy difference between the input and output regions. In addition, we assume time-independent potentials in the chiral region, under which electron energy is conserved. Here, we extend the analysis to two additional scenarios. First, we introduce a finite bias by increasing (decreasing) the potential energy of the input (output) region. Second, we consider time-dependent potentials within the chiral region, so that electron energy is no longer conserved. In both cases, we find that the CIOS effect remains robust.

In Fig.~\ref{figS10}(a), the potential energy of the input (output) region is increased (decreased) by $V_3/2$, resulting in a finite bias of $V_3$. The time-independent Gaussian potentials with $L=2\pi P$, used in Fig.~3 of the main text, are assumed in the chiral region. For the two representative cases $V_1\in\{0.1,0.4\}\,{\rm eV}$ and $V_2=0.2\,{\rm eV}$ considered in Fig.~3 of the main text, Fig.~\ref{figS10}(a) shows the transmittance difference $\Delta T$ as a function of the finite bias $V_3$. The initial kinetic energy of the electron wave packet is taken to be ${\rm KE}_0=1\,{\rm eV}$, so that the total energy becomes ${\rm KE}_0+V_3/2$. Notably, the transmittance difference $\Delta T$ remains on the order of $0.1$ over a broad range of the bias, $V_3\in [0,2]\,{\rm eV}$, indicating that the CIOS effect is robust in the presence of the finite bias.

In Figs.~\ref{figS10}(b)-(d), the zero-bias case ($V_3=0$) is considered with time-dependent Gaussian potentials in the chiral region. The offset is taken as $V_1\in\{0.1,0.4\}\,{\rm eV}$, with the corresponding results shown in red and blue, respectively. The amplitude of the Gaussian potential centered at position $z$ is assumed to be time-dependent and modeled as
$V_2 + \sum_{j=1}^{3} A_j \sin(k_j z + \omega_j t + \phi_j)$ with $V_2=0.2\,{\rm eV}$. For each simulation, the amplitudes $A_j$ of the fluctuating terms are randomly drawn from independent normal distributions with zero mean and standard deviation $\sigma_A$. Similarly, the phases $\phi_j$ are randomly drawn from uniform distributions over $[0,2\pi]$. The wave vectors are taken as $k_j = j/P = 2\pi j/L$, with dispersion relations $\omega_j = \mu k_j$, where $\mu$ sets the time scale of the fluctuating potentials. Figures~\ref{figS10}(b)-(d) show statistical data obtained from simulations with the randomly generated amplitudes $A_j$ and phases $\phi_j$.

In Fig.~\ref{figS10}(b), the distributions of the kinetic energy of the electron wave packet in the output region, after scattering within the chiral region, are shown as a function of the parameter $\mu$ in the dispersion relations, with fixed $\sigma_A=0.2\,{\rm eV}$. For a detailed kinetic energy analysis, here we consider a one-dimensional model without orbital degrees of freedom. Note that, due to the time-dependent potentials in the chiral region, the average kinetic energy of the output electron wave packet differs from the initial kinetic energy ${\rm KE}_0 = 1\,{\rm eV}$. We consider $\mu\in\{0.2,0.5\}\,{\rm nm/fs}$ as two representative cases, and provide further analysis in Figs.~\ref{figS10}(c) and (d), respectively.

In Fig.~\ref{figS10}(c), for $\mu = 0.2\,{\rm nm/fs}$, the kinetic energy of the output electron is shown as a function of $\sigma_A$ (top panel, 1D model), indicating that electron energy is not conserved over the $\sigma_A$ range considered in the simulations. For the 3D model including orbital degrees of freedom, the transmittance difference $\Delta T$ (middle panel) and orbital polarization (bottom panel) are shown as functions of $\sigma_A$, demonstrating that the CIOS effect remains robust under the time-dependent potentials. Similar results are obtained for $\mu = 0.5\,{\rm nm/fs}$, as shown in Fig.~\ref{figS10}(d).

\section{Orbital to Spin Selectivity}

\begin{figure}
    \includegraphics[width=0.5\linewidth]{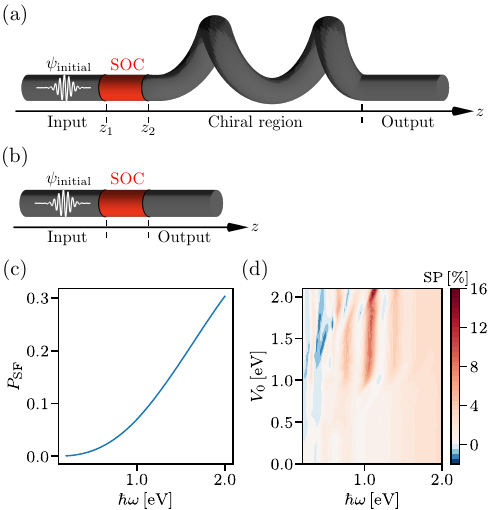}
    \caption{(a) Schematic representation of a 3D model for electron transfer through input-chiral-output regions, where SOC is present in the input region for $z_1 < z < z_2$. (b) Simplified setup without the chiral region. (c) Population of spin-flipped states transmitted through the SOC region, when the chiral region is omitted (see (b)). (d) Spin polarization of the full model in the presence of the chiral region with zero SOC (see (a)).}
    \label{figS11}
\end{figure}

In the main text, we discuss that orbital selectivity can give rise to spin selectivity when spin-OAM states are correlated before transmission to the chiral region, even if the spin-orbit interaction within the chiral region is negligible. Here we provide an example by considering a finite input region with spin-orbit coupling (SOC) for $z_1\le z\le z_2$, as schematically shown in Fig.~\ref{figS11}(a). The SOC is modeled by \begin{equation}
    H_{\rm{SOC}} =\alpha_0 \boldsymbol{\sigma}\cdot(\nabla W\times\mathbf{p}) \Theta(z-z_1)\Theta(z_2-z),
\end{equation}
where $W$ is the harmonic potential energy within the achiral input region, where equilibrium positions are independent of $z$, and $\Theta(z)$ is the step function defined as $\Theta(z)=1$ for $z>0$ and $\Theta(z)=0$ otherwise. For simplicity, we assume that the length of the SOC region is $z_2-z_1=1\,\rm{nm}$ and $\hbar \alpha_0=0.01\,\rm{nm^2}$, so that the spin-energy splitting scales as $\hbar\alpha_0m\omega^2\in\{0.005,0.5\}\,\rm{eV}$ for $\hbar\omega\in\{0.2,2\}\,\rm{eV}$. In the simulations, we consider a random mixture of spin up and down states at the initial time, where an electron is localized in the input region without SOC (i.e., $z<z_1$) and carries zero OAM but has linear momentum along the $z$-direction, with the initial kinetic energy of ${\rm KE}_0=1\,{\rm eV}+\hbar\omega$. The additional energy $\hbar\omega$ is introduced so that, when the OAM becomes $\hbar$ or $-\hbar$ via the SOC, the kinetic energy along the propagation $z$-direction is reduced to $1\,{\rm eV}$, consistent with the simulation settings in the main text.

As the electron wave packet propagates through the SOC region, the population of the initial angular momentum states, namely, $\ket{n_c=0,n_d=0,\uparrow}$ or $\ket{n_c=0,n_d=0,\downarrow}$, is converted into that of the spin-flipped angular momentum states, $\ket{n_c=1,n_d=0,\downarrow}$ or $\ket{n_c=0,n_d=1,\uparrow}$, while conserving the $z$-component of total angular momentum. Here, the net spin polarization is zero as the SOC region is achiral. However, the spin-OAM states are correlated, as $\ket{n_c=0,n_d=0,\uparrow}$ and $\ket{n_c=1,n_d=0,\downarrow}$ (or $\ket{n_c=0,n_d=0,\downarrow}$ and $\ket{n_c=0,n_d=1,\uparrow}$) are superposed via the SOC when the initial state carries spin up (or down). The other states carrying the same total angular momentum $\hbar/2$ (or $-\hbar/2$), such as $\{\ket{n_c=N,n_d=N,\uparrow},\ket{n_c=N+1,n_d=N,\downarrow}\}$ (or $\{\ket{n_c=N,n_d=N+1,\uparrow},\ket{n_c=N,n_d=N,\downarrow}\}$) with positive integer $N$, have negligible contributions to the electron dynamics in the input region, as the SOC considered in our simulations is not strong enough to significantly populate these states. The total angular momentum conservation arises as the Hamiltonian of the achiral input region is invariant under rotations about the $z$-axis, ensuring that the $z$-component of the total angular momentum remains constant until the electron enters the chiral region.

To quantify the population converted into the spin-flipped states, we omit the chiral region, as shown in Fig.~\ref{figS11}(b), and examine the spin-flipped population $P_{\rm SF}$ as a function of $\omega$, as shown in Fig.~\ref{figS11}(c). When $\hbar\omega\approx 1\,{\rm eV}$, the spin-flipped population is approximately $0.07$. Although the spin-flipped population, along with the corresponding spin-OAM correlations, is small, it can nevertheless give rise to a notable spin selectivity. In Fig.~\ref{figS11}(d), we consider the full setup, including the chiral region without SOC (see Fig.~\ref{figS11}(a)) and a rectangular energy barrier of height $V_0$ located after the chiral region, and show the spin polarization (SP) of the transmitted electron wave. Notably, an SP of the order of $10\,\%$ is observed at $\hbar\omega\approx 1\,{\rm eV}$. These results demonstrate that even if the chiral region has negligibly weak SOC, the orbital selectivity of the chiral region can give rise to spin selectivity when the spin-OAM states are correlated prior to entering the chiral region, e.g., via relatively strong spin-orbit interaction within metallic electrodes.

\section{Numerical Methods and Implementation}

To implement numerical simulations of the three-dimensional continuous-variable models considered in this work, we discretized the $z$ coordinate and approximated derivatives with respect to $z$, appearing in the momentum and kinetic operators, using a fourth-order finite-difference method. The grid spacing in $z$ was systematically reduced until convergence of the simulation results was achieved. To reduce the total number of discretized $z$ points, we introduced absorptive layers at the ends of the input and output regions. These were implemented using an imaginary term added to the Hamiltonian
\begin{align}
    \Gamma_{a}(z) &= -i\gamma_a (1-e^{-(z-z_{0})^2/2\sigma^2_{a}})/2,
\end{align}
where $z_0$ denotes the $z$-coordinate of the open end of the input or output region. The parameters $\gamma_a$ and $\sigma_{a}$ were chosen such that dissipation of the electron wave packet by the imaginary term is negligible until the initial electron wave packet is scattered from the chiral region, while wave packets reflected from or transmitted through the chiral region are efficiently absorbed, ensuring that re-entry of these wave packets into the chiral region is negligible. For example, in the simulations shown in Fig.~3, we used 40 discretized $z$ points per nm, corresponding to a total of 3,736 discretized $z$ points over the input, chiral, and output regions. For the transverse motion of the electron in the $xy$-plane, described in the eigenbasis $\ket{n_c,n_d}$ of the harmonic oscillators, we considered 10 levels in each direction, namely $n_c,n_d\in\{0,1,\cdots,9\}$, to obtain numerical convergence of the simulation results in Fig.~3, leading to a total dimension of 373,600. Comparable or even higher dimensions were considered in the simulations shown in other figures.

We computed the time evolution of the electron wave packet using a fourth-order Runge-Kutta method with a time step of $dt=0.002\,\rm{fs}$ to achieve convergence of the simulation results. The transmittance $T_{n_c,n_d}$ was obtained by computing the population dissipated by the absorptive layer in the output region.

The lattice-like potentials considered in Fig.~3 and Fig.~\ref{figS10} were multiplied by a smooth step function to ensure a smooth transition between input, chiral and output regions, defined as
\[
f(z;z_0,z_1) = 
\begin{cases}
    192(z-z_0+0.5)^5-240(z-z_0+0.5)^4+80(z-z_0+0.5)^3, & z_0-0.5<z<z_0 \\
    1, & z_0\leq z\leq z_1 \\
    1-(192(z-z_1)^5-240(z-z_1)^4+80(z-z_1)^3), & z_1<z<z_1+0.5
\end{cases}
\]
for a chiral region $z_0<z<z_1$.

The static disorder in Fig.~3(d) and dynamic disorder in Fig.~\ref{figS10} were implemented by running multiple simulations with randomly sampled parameters, followed by ensemble averaging. For each case, we used 200 and 1000 realizations, respectively.

\end{widetext}
\end{document}